\documentclass[preprint]{revtex4-1}
\usepackage[cp1251]{inputenc}
\usepackage{amsmath} 
\usepackage{amsfonts} 
\usepackage{graphicx}
\usepackage[titletoc,toc,title]{appendix}
\usepackage{hyperref}   
\usepackage{subfigure}
\usepackage{float}

\newcommand{\n}{\\ \nonumber}
\newtheorem{theorem}{Proposition}

\begin{document}
\title{Convergent series for lattice models with polynomial interactions}

\author{Aleksandr S. Ivanov}
\affiliation{M. V. Lomonosov Moscow State University, Faculty of Physics, Leninskie Gory, 119991, Moscow, Russia}
\affiliation{Institute for Nuclear Research RAS, 60-letiya Oktyabrya prospekt 7a, 117312, Moscow, Russia}
\email{ivanov.as@physics.msu.ru}
\author{Vasily K. Sazonov}
\affiliation{Institut f\"{u}r Physik, FB Theoretische Physik, Universit\"{a}t Graz, Universit\"{a}tsplatz 5, A-8010, Graz, Austria}
\email{vasily.sazonov@uni-graz.at}
        
\begin{abstract}
The standard perturbative weak-coupling expansions in lattice models
are asymptotic. The reason for this is hidden in the
incorrect interchange of the summation and integration. 
However, substituting the Gaussian initial approximation of the perturbative expansions by a certain interacting model
or regularizing original lattice integrals, one can construct desired convergent series.
In this paper we develop methods, which are based on the joint and separate utilization of the regularization and
new initial approximation. We prove, that the convergent series exist and
can be expressed as the re-summed standard perturbation theory for any model
on the finite lattice with the polynomial interaction of even degree.
We discuss properties of such series and make them applicable to practical
computations. The workability of the methods is demonstrated on the example of the lattice $\phi^4$-model.
We calculate the operator $\langle\phi_n^2\rangle$ using the convergent series,
the comparison of the results with the Borel re-summation and Monte Carlo simulations
shows a good agreement between all these methods.
\end{abstract}

\maketitle

\section{Introduction}
The main objects of studies in quantum field theory are the Green's functions. They can be naturally
expressed in terms of the path integral. However, the path integral itself
gives only a formal solution and for the practical computations one has to find 
some efficient method of its evaluation. One of the possibility is to use a finite lattice approximation
of the physical space and then investigate the truncated system using the Monte Carlo
method. Such approach was successfully applied in many areas of statistical, condensed matter 
and elementary particle physics for a wide range of physical parameters, {\it e.g.} coupling constants. 
The Monte Carlo method is based on the probabilistic
interpretation of the lattice integrals measure and, consequently, it can not be applied, when the action of the
system is complex. For instance, this complex action problem, or in other words sign problem, impedes the 
the computations in lattice QCD at finite chemical potentials \cite{deForcrand}.
Another traditional way of the path and lattice integrals computations is the utilization of the 
standard perturbation theory with the Gaussian initial approximation (SPT). 
Usually, the series of SPT are asymptotic \cite{Dyson, Lipatov} and applicable 
only for the small coupling constants (parameters of the expansion). For example, 
the most precise results within the standard perturbation theory were obtained in QED \cite{Laporta1996, Aoyama2012}, 
where the expansion parameter is extremely small: $\alpha \simeq 1/137$. 
At the same time SPT fails to describe the low energy physics of graphene, which is effectively
determined by QED with large coupling constant \cite{Kolomeisky2015}.

The SPT series are asymptotic because of the incorrect interchange of the summation and
integration during the construction of the perturbative expansion. 
Nevertheless, the problem of the interchanging can be avoided. 
For instance, it can be done by an appropriate regularization of the original integral. The regularization
by cutting off the large fluctuations of the fields in the lattice models was suggested in \cite{Meurice2004, Meurice20052}.
A more sophisticated and delicate regularization method providing
the construction of the convergent perturbation theory for the lattice models and path integrals
with the measure defined by the trace-class operators was developed in works 
\cite{Belokurov1, Belokurov2, Belokurov3, BelSolSha97, BelSolSha99}.
The method, based on the modification of the interaction power due to the application of the intermediate field representation
and on the consequent utilization of the forests/trees formula \cite{Brydges, TreesForests} was developed in works
\cite{tears, ZeroDimLVE, phi2k, HowToResum, correctedPhi42}.
Alas, the computations with all these methods are highly complicated.

An alternative approach to the construction of the convergent perturbative series in scalar field theories,
based on changing of the initial Gaussian approximation to a certain interacting theory,
was proposed in \cite{Ushveridze1983, Shaverdyan1983, UshveridzeSuper}. Later, independently developed and similar ideas become
a basis of the variational perturbation theory methods \cite{Feynman}.
In \cite{Turbiner} the approach \cite{Ushveridze1983, Shaverdyan1983, UshveridzeSuper} was extended
to the construction of the strong coupling expansion for the anharmonic oscillator.
In \cite{Nalimov} the RG-equations consistent with the method \cite{Ushveridze1983} were derived.
The critical indices of the $\varphi^4$-model obtained within the latter RG-equations
are in close agreement with the experimental and numerical results. However, a rigorous
mathematical proof of the convergence of expansions \cite{Ushveridze1983, Shaverdyan1983, UshveridzeSuper, Nalimov}
is still missing. 
A non detailed version of the rigorous construction of the convergent 
series similar to \cite{Shaverdyan1983, UshveridzeSuper} 
for the one-dimensional lattice $\phi^4$-model was presented in \cite{IvanovProc}. 
The numerical computations within the convergent series in \cite{IvanovProc}
revealed a perfect agreement with the results obtained by the Monte Carlo simulations for the lattice
with the volume $V = 2$ and demonstrated slow convergence to the correct answers even for 
the slightly bigger lattices (with $V = 4$ and $V = 8$ lattice sites). 

Here we continue the investigation of the convergent series method, carrying out the main derivations on the example of the lattice $\phi^4$-model
and performing generalizations, when it is necessary. 
We prove, that the convergent series (CS) can be rigorously constructed 
for any finite lattice model with the polynomial interaction of even degree 
and that the series is expressed as a re-summation of SPT. We show, 
that up to all loops of the standard perturbation theory the convergent series method
has an internal symmetry, providing the possibility to introduce a variational parameter.
The freedom in a choice of the variational parameter allows one to obtain numerical results,
which are in agreement with the Monte Carlo data.
Though, the convergence and non-perturbative correctness of CS, modified by the the variational
parameter, is not {\it a priori} evident. In the following we denote the latter series as variational (VS).
To investigate the convergence of the variational series, 
we consider two regularizations of the lattice $\phi^4$.
The first one, $\eta$-regularization, is a natural extension of the variational series construction and it
gives suggestive arguments about the convergence of VS.
The second, $\gamma$-regularization, is also based on the mathematical structures, used in VS.
We proof, that $\gamma$-regularized model approximates original $\phi^4$-model with any arbitrary
precision and that the Green's functions of the $\gamma$-regularized model can be calculated with
the variational series, which is convergent in this case.
We demonstrate the non-perturbative independence on the variational parameter of VS, when the 
$\gamma$-regularization is removed. Using this independence property, we propose a way for the computations
in the infinite volume limit.
Summarizing the properties of CS and VS with regularizations, we conjecture the convergence of the variational series.
We study the applicability of the CS and VS for different lattice volumes 
and investigate the dependence on the variational and regularization
parameters, computing the operator $\langle\phi_n^2\rangle$ of the lattice $\phi^4$-model.
The results are compared with the Monte Carlo simulations and Borel re-summation.

The paper is organized as follows. In Section \ref{constr} the convergent series
for lattice models with polynomial interactions is constructed.
We introduce variational series and discuss its perturbative properties in Section \ref{impr}.
To investigate non-perturbative aspects of VS we study regularized lattice models in Section \ref{conv}.
The results of the numerical computations are presented in Section \ref{numer}.
We conclude in Section \ref{concl}.

\section{Construction of the convergent series}
\label{constr}

We start with the construction of the convergent series for the lattice $\phi^4$-model. The model is defined by the action
\begin{eqnarray}
S[\phi] = \frac{1}{2} \sum_{m,n = 0}^{V-1} \phi_m K_{mn} \phi_n + \frac{\lambda}{4!} \sum_{n = 0} ^ {V-1} \phi _ {n} ^ 4,
\label{Sphi4}
\end{eqnarray}
\begin{eqnarray}
\frac{1}{2} \sum_{m,n = 0}^{V-1} \phi_m K_{mn} \phi_n = 
\sum_{n = 0}^{V - 1}\Big[\frac{1}{2} M^2 \phi_{n}^2
-\frac{1}{2}\sum_{\mu = 1}^d \big(\phi_{n + \hat\mu} + \phi_{n - \hat\mu} - 2 \phi_{n}\big)\phi_{n}\Big]\,,
\label{Kmn}
\end{eqnarray}
where $M$ is a mass parameter, $\lambda$ is a coupling constant, $V$ is the volume of the lattice,
indices $m$ and $n$ label the lattice sites, $d = 1,2$ is a dimension of the lattice, 
index $\mu$ runs over all spatial dimensions and $\hat\mu$ stands for the unit vector in the corresponding direction.
The periodical boundary conditions in all possible directions are assumed.

Without loss of generality, as an example of an arbitrary Green's function, we consider the two point Green's function (propagator). 
The normalized to the free theory propagator is defined as
\begin{eqnarray}
  \langle\phi_i \phi_j\rangle = \int\, [d\phi]\, \phi_i \phi_j\, \exp\{-S[\phi]\}\,,
\label{correl}
\end{eqnarray}
where $\int [d\phi] = \frac{1}{Z_0} \prod_{n} \int\, d\phi_n$ and
\begin{eqnarray}
  Z_0 = \prod_{n} \int\, d\phi_n\, e^{-\frac{1}{2} \sum_{l,m} \phi_l K_{lm} \phi_m}
\end{eqnarray}
is the partition function of the free theory.
Following \cite{Shaverdyan1983, UshveridzeSuper}, we split the action into the
new non-perturbed part $N[\phi]$ and perturbation: $S[\phi] = N[\phi] + (S[\phi] - N[\phi])$.
For the following it is convenient to modify the latter expression as
\begin{eqnarray}
  S_\eta[\phi] = N[\phi] + \eta (S[\phi] - N[\phi])\,,
\label{split}
\end{eqnarray}
where $\eta \leq 1$ is a parameter labeling the order of the new perturbative expansion.
The parameter $\eta$ will be also used in Section \ref{etaReg} for the regularization.
In the current section we derive formulas containing $\eta$, keeping in mind, that the initial model corresponds to $\eta = 1$. 
Then, the propagator (\ref{correl}) can be written as
\begin{eqnarray}
  \langle\phi_i \phi_j\rangle = \int\, [d\phi]\, \phi_i \phi_j\, e^{-N[\phi]}
  \sum_{l = 0}^\infty
  \frac{\eta^l (N[\phi] - S[\phi])^l}{l!}\,.
\label{correl2}
\end{eqnarray}
When
\begin{eqnarray}
  N[\phi] \geq S[\phi]\,,
\label{NgS}
\end{eqnarray} 
the interchanging of the summation and integration in (\ref{correl2}) 
leads to an absolutely convergent series
\begin{eqnarray}
  \langle\phi_i \phi_j\rangle =  \sum_{l = 0}^\infty \langle\phi_i \phi_j\rangle_l
\label{sum}
\end{eqnarray}
with terms given by
\begin{eqnarray}
  \langle\phi_i \phi_j\rangle_l = \frac{\eta^l}{l!} \int\, [d\phi]\, \phi_i \phi_j\, \big(N[\phi] - S[\phi]\big)^l\,e^{-N[\phi]}\,.
\label{lf}
\end{eqnarray}
Indeed, it is easy to see, that
\begin{eqnarray}
\nonumber
  \Big| \sum_{l = 0}^\infty \langle\phi_i \phi_j\rangle_l\Big| \leq  \sum_{l = 0}^\infty |\langle\phi_i \phi_j\rangle_l|\, \leq
   \sum_{l = 0}^\infty \frac{\eta^l}{l!} \int\, [d\phi]\, |\phi_i \phi_j|\,
\big(N[\phi] - S[\phi]\big)^l\,e^{-N[\phi]}\\
  = \int\, [d\phi]\, |\phi_i \phi_j|\, e^{-S_\eta[\phi]} < \infty\,,~~\text{for}~~ \eta \leq 1\,.
\label{est10}
\end{eqnarray}
There are many possibilities to choose the new initial approximation $N[\phi]$, satisfying inequality~(\ref{NgS}),
however it should correspond to a solvable model.
Here we take
\begin{eqnarray}
  N[\phi] = \sum_{n, m} \frac{1}{2} \phi_n K_{n m} \phi_m + \sigma \Big(\sum_{n, m} \frac{1}{2} \phi_n K_{n m} \phi_m\Big)^2\,,
\label{N}
\end{eqnarray}
where $\sigma$ is an unknown positive parameter, which is determined by the 
substitution of (\ref{N}) into (\ref{NgS}):
\begin{eqnarray}
  \sigma \geq \frac{\lambda}{6 M^4}\,.
\end{eqnarray}

The functions (\ref{lf}) can be calculated in the following way. Introducing an
auxiliary integration, we change $\lVert\phi\rVert \equiv \Big(\frac{1}{2} \phi_n K_{n m} \phi_m\Big)^{\frac{1}{2}}$
to the one-dimensional variable $t$
\begin{eqnarray}
  \langle\phi_i \phi_j\rangle_l = \frac{\eta^l}{l!} 
  \int\, [d\phi]\, 
  \phi_i \phi_j\, \int_{0}^\infty dt\, e^{-t^2 - \sigma t^4}\,\delta(t - \lVert\phi\rVert) 
\Big(\sigma t^4 - \frac{\lambda}{4!} \sum_n \phi_n^4 \Big)^l\,.
\label{t0}
\end{eqnarray}
Rescaling the field variables as $\phi_n^{old} = t \phi_n$, we get
\begin{eqnarray}
  \langle\phi_i \phi_j\rangle_l =  J_\eta(V, l) \int\, [d\phi]\, \phi_i \phi_j\,\delta(1 - \lVert\phi\rVert) 
  \Big(\sigma - \frac{\lambda}{4!} \sum_n \phi_n^4 \Big)^l\,,
\label{t1}
\end{eqnarray}
where 
\begin{equation}
  J_\eta(V, l) = \frac{\eta^l}{l!} \int_{0}^\infty dt\, e^{-t^2 - \sigma t^4} t^{V + 4l + 1}\,.
\end{equation}
The factor $t^{V + 4l - 1}$ in the integrand in $J_\eta(V, l)$ is obtained by rescaling of the fields measure, fields and
delta function.
Now the multi-dimensional (lattice) part of the integral is factorized from the auxiliary integration.
Applying the binomial expansion to the brackets $(...)^l$, we rewrite (\ref{t1}) as
\begin{eqnarray}
  \langle\phi_i \phi_j\rangle_l = J_\eta(V, l)
  \int\, [d\phi]\, \phi_i \phi_j\,\delta(1 - \lVert\phi\rVert) 
  \sum_{k = 0}^{l} C_l^k \sigma^{l - k} \Big(-\frac{\lambda}{4!} \sum_n \phi_n^4 \Big)^k\,.
\label{fle}
\end{eqnarray}
To solve (\ref{fle}), we use the following equality
\footnote{The r.h.s. of the identity (\ref{ident}) is obtained from the left one by the transformations analogous to (\ref{t0}, \ref{t1}).}
\begin{eqnarray}
  \int\, [d\phi]\, \phi_{n_1}...\phi_{n_Q}\,e^{-\lVert\phi\rVert^2} = 
  \frac{1}{2}\Gamma\Big(\frac{V + Q}{2}\Big) \int\, [d\phi]\, \phi_{n_1}...\phi_{n_Q}\, \delta(1 - \lVert\phi\rVert) \,.
\label{ident}
\end{eqnarray}
By substituting (\ref{ident}) in (\ref{fle}), we obtain
\begin{eqnarray}
  \langle\phi_i \phi_j\rangle_l  = J_\eta(V, l) 
  \sum_{k = 0}^{l} C_l^k \sigma^{l - k} \frac{2}{\Gamma\Big(\frac{V + 4 k + 2}{2}\Big)}
  \int\, [d\phi]\, \phi_i \phi_j\,
  e^{-\lVert\phi\rVert^2}
  \Big(-\frac{\lambda}{4!} \sum_n \phi_n^4 \Big)^k\,.
\label{main0}
\end{eqnarray}
Denoting the $k$-th order of the standard perturbation theory as $f_k$, we rewrite (\ref{main0}) as
\begin{eqnarray}
  \langle\phi_i \phi_j\rangle_l  = J_\eta(V, l)
  \Bigg[\sum_{k = 0}^{l} C_l^k \sigma^{l - k} \frac{2\,k!\, f_k}{\Gamma\Big(\frac{V + 4 k + 2}{2}\Big)}\Bigg]\,.
\label{main}
\end{eqnarray}
Therefore, each certain order $l$ of the convergent series is expressed as a linear combination of first $l$ 
orders of SPT
with coefficients given by the one-dimensional analytically calculable $t$-depending integrals.
However, the latter fact does not mean that the convergent series looses non-perturbative contributions
from non-analytical functions such as $e^{-\frac{1}{\lambda}}$. Being an expansion with the non-Gaussian initial
approximation, it automatically takes non-perturbative contributions into account in a similar way,
as the function $e^{-\frac{1}{\lambda}}$ for positive $\lambda$ can be reproduced by its Taylor series around $\lambda = 1$.

The lattice $\phi^4$ is a Borel summable model, however, it is possible to generalize the results of the current section
to a wider class of models, which include Borel non-summable cases.
\begin{theorem}
Consider a model on the finite lattice, determined by the polynomial action $S[\phi] = P[\phi]$ with
an even $deg(P)$, then it is always possible to construct a convergent series for this model
with the terms which can be expressed as a linear combinations of terms of the standard perturbation theory.
\end{theorem}
The proof follows from the construction presented above and from the fact, that each polynomial
can be bounded as
\begin{equation}
  |P[\phi]| \leq const \,(1 + \lVert\phi\rVert^{deg(P)})\,.
\end{equation}
Moreover, it is possible to show, that convergent series is a re-summation of the standard perturbation theory,
for details see Section \ref{impr}.

The latter proposition demonstrates, that the whole non-perturbative physics of the lattice models with the polynomial actions
can be encoded by the coefficients of the standard perturbation theory. However, it is important to note, that this
does not mean, that the non-perturbative information can be obtained only from the standard perturbative expansion.
An additional input, needed for the construction of the convergent series is received from the model itself.
In some sense it is similar to the resurgence program \cite{Dorigoni}, where for the recovering of the non-perturbative contributions
from the perturbative series, it is assumed that the solution is a resurgent function.

\section{Variational series}
\label{impr}
The previous studies of the convergent series application to the lattice $\phi^4$-model 
\cite{IvanovProc} demonstrated a critical slow down of the convergence rate with the increasing of the lattice volume $V$
(see Section \ref{numer}).
However, this problem can be resolved by the following observation.
When $\eta = 1$, the explicit dependence on the lattice volume $V$ in the sum (\ref{sum}) of functions (\ref{main})
can be substituted by $\tau = V + \alpha$, {\it i.e.},
\begin{eqnarray}
  \langle\phi_i \phi_j\rangle = 
  \sum_{l = 0}^\infty J_{\eta = 1}(\tau, l)
  \Bigg[\sum_{k = 0}^{l} C_l^k \sigma^{l - k} \frac{2\, k!\, f_k}{\Gamma\Big(\frac{\tau + 4 k + 2}{2}\Big)}\Bigg]\,,~
\label{main2o}
\end{eqnarray}
where $\tau > -2$, not to generate singularities in the integrals $J_\eta(\tau, l)$.
The latter inequality is special for the propagator, for an arbitrary $n$-fields Green's function
it has to be substituted by $\tau > -n$.

Let us first consider only a perturbative proof of (\ref{main2o}) (the non-perturbative aspects are considered in the Section \ref{conv}). 
It can be obtained by changing the order of summations in~(\ref{main2o})
\begin{eqnarray}
  \langle\phi_i \phi_j\rangle \approx \sum_{k = 0}^\infty\sum_{l = 0}^\infty J_{\eta = 1}(\tau, l)
   C_l^k \sigma^{l - k} \frac{2\, k!\, f_k}{\Gamma\Big(\frac{\tau + 4 k + 2}{2}\Big)}
   \,,
\label{proof1}
\end{eqnarray}
where the sign '$\approx$' stands to indicate only perturbative equivalence between left and right parts of the expression.
The summands in the latter expression are equal to zero, when $l < k$. Changing the summation index to $y = l - k$,
we obtain
\begin{eqnarray}
\nonumber
  \langle\phi_i \phi_j\rangle \approx \sum_{k = 0}^\infty \frac{2\,f_k\, \int_{0}^\infty dt\, t^{\tau + 4 k + 1} e^{-t^2 - \sigma t^4}
  \sum_{y = 0}^\infty \frac{C_{(y+k)}^k\, k!\, \sigma^{y} t^{4y}}{(y + k)!}}
  {\Gamma\Big(\frac{\tau + 4 k + 2}{2}\Big)} \n
  = \sum_{k = 0}^\infty \frac{2\,f_k\, \int_{0}^\infty dt\, t^{\tau + 4 k + 1} e^{-t^2 - \sigma t^4}
	  \sum_{y = 0}^\infty 
	  \frac{
		\sigma^{y} t^{4y}}{y!
	  }
   }{\Gamma\Big(\frac{\tau + 4 k + 2}{2}\Big)} 
  \\
  = \sum_{k = 0}^\infty \frac{2\,f_k\, \int_{0}^\infty dt\, t^{\tau + 4 k + 1}
 e^{-t^2}}{\Gamma\Big(\frac{\tau + 4 k + 2}{2}\Big)} = \sum_{k = 0}^\infty f_k\,.
\label{proof2}
\end{eqnarray}
We end up with the series of the standard perturbation theory.
There are two important consequences from this fact (the non-perturbative analogues of these statements are derived in the following sections).
\begin{itemize}
  \item The whole sum of the series over $l$ in (\ref{main2o}) does not depend on $\tau$.
	Therefore, $\tau$ is a variational parameter and can be taken arbitrary to optimize the convergence of the series.
  \item According to (\ref{proof2}), the convergent series is a re-summation method. Thus,
	for the computation of the connected Green's functions (including the normalized to the full partition sum propagator,
	which is a subject of our numerical studies) one can use the fact, that in the standard perturbation theory
	connected functions are obtained from the full Green's functions by throwing away disconnected Feynman diagrams from the expansions.
\end{itemize}

\section{Convergence of the variational series}
\label{conv}
In the last section we have introduced a variational parameter $\tau$, to improve the convergence
rate of CS (see Section \ref{numer}). However, the derivations of the Section \ref{constr}
are not applicable when $\tau \neq V$. The proof of the convergence of the series (\ref{sum}) is based on the positivity
of the brackets $(N[\phi] - S[\phi])^l$. When $\tau \neq V$, each summand of the binomial expansion of
$(N[\phi] - S[\phi])^l$ transforms differently under the change of $\tau$ and
the positivity of the brackets $(N[\phi] - S[\phi])^l$ can be lost. For instance, it is the case
for the one-site lattice $\phi^4$-integral at $\tau = 0$.
In Section \ref{impr} we have proved the perturbative independence on $\tau$ of the total sum of the series (\ref{main2o}).
One can not {\it a priori} exclude the possibility, that some non-analytical dependence on $\tau$ at $\eta = 1$, which gives zero contribution to the SPT series, still
persist in~(\ref{main2o}). The prescription for the evaluation of the connected Green's functions, suggested in the previous section, is based also
only on the standard perturbation theory arguments. 

Here we study all these issues from the non-perturbative point of view.
For this we investigate the convergence properties of the variational expansions constructed for the lattice $\phi^4$
with two different regularizations. The first one is the $\eta$-regularization and it is achieved by considering $0 < \eta < 1$
in (\ref{split}). The second, $\gamma$-regularization, is defined in Section~\ref{gammaReg} by introducing
an additional term proportional to $\|\phi\|^6$ into the action (\ref{Sphi4}).

\subsection{Convergence of the variational series depending on $\eta$ and $\tau$}
\label{etaReg}
\begin{figure}[H]
  \begin{center}
      \includegraphics[width=0.5\linewidth, angle = 0]{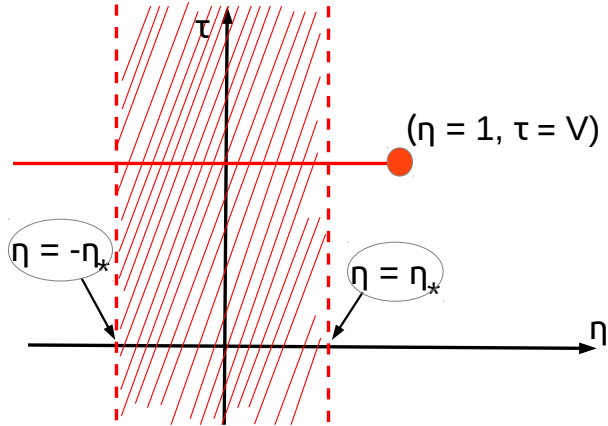}
      \caption{The convergence of the series (\ref{main2}) in the band from $-\eta_*$ to $\eta_*$ is provided by the bound~(\ref{main3}).
      The convergence at $\tau = V$ and $\eta \leq 1$ follows from the estimate (\ref{est10}).}
      \label{epsalph}
  \end{center}
\end{figure}

Let us study the convergence of the series for the full propagator, obtained from (\ref{main2o}), extending it to $\eta \leq 1$
\begin{eqnarray}
  \langle\phi_i \phi_j\rangle = 
  \sum_{l = 0}^\infty J_{\eta}(\tau, l)
  \Bigg[\sum_{k = 0}^{l} C_l^k \sigma^{l - k} \frac{2\, k!\, f_k}{\Gamma\Big(\frac{\tau + 4 k + 2}{2}\Big)}\Bigg]\,.
\label{main2}
\end{eqnarray}
The asymptotic of large orders of the perturbation theory
in quantum field theories and lattice models for the connected and full correlation functions
have similar form \cite{Lipatov, Spencer, ParisiScQED}
\begin{equation}
  f_k\, \sim\, (-1)^k\, \sqrt{2\pi}\, e\, \Big(\frac{a}{e}\Big)^k\, k^{k +b_0 + 1/2}\,,
\label{asymp2}
\end{equation}
where $a, b_0 \in \mathbb{R}$ are some constants and $e$ is the Euler's number.
The upper bound for the series
(\ref{main2}) can be obtained as
\begin{eqnarray}
\nonumber
  |\langle\phi_i \phi_j\rangle|
  \leq
  \sum_{l = 0}^\infty \big|J_\eta(\tau, l)\big|
  \Big[\sum_{k = 0}^{l} C_l^k \sigma^{l - k} \frac{2\, k!\, |f_k|}{\Gamma\Big(\frac{\tau + 4 k + 2}{2}\Big)}\Big] \\
  = \sum_{k = 0}^\infty \frac{2 \int_{0}^\infty dt\, t^{\tau + 4 k + 1} e^{-t^2 - \sigma (1 - |\eta|) t^4}}{\Gamma\Big(\frac{\tau + 4 k + 2}{2}\Big)}
  |f_k| |\eta|^k
  \,.~~~~
\label{main3}
\end{eqnarray}
The coefficients in front of $|f_k|$ in (\ref{main3}) at large $k$ behave as
\begin{equation}
%
  2^{1/2 - 2\beta} \sigma_\eta^{-\beta} k^{-k} \Big(\frac{e |\eta|}{4 \sigma_\eta}\Big)^k\,,
\label{asymp3}
\end{equation}
where $\beta = (\tau + 2) / 4$ and $\sigma_\eta = \sigma (1 - |\eta|)$.
Consequently, the bound (\ref{main3}) is convergent, when
$|\eta| < \eta_* = \frac{4 \sigma}{|a| + 4\sigma}$, independently on the value of $\tau$. 
When $\tau = V$, the series (\ref{main2}) converges for $\eta \leq 1$ due to the estimate (\ref{est10}). 
The bound
(\ref{main3}) has finite radius of the convergence in terms of $\eta$ for any $\tau > -2$,
including, for instance $\tau = 0$. 
In Fig. \ref{epsalph} we show the area of the parameters $\tau$ and $\eta$ for which the convergence
of the series (\ref{main2}) is guaranteed by (\ref{main3}) and~(\ref{est10}).

The convergence of the series (\ref{main2}) is better
than the convergence of (\ref{main3}), because of the cancellations in the internal sums over $k$ in (\ref{main2}).
Therefore, it might be, that the series (\ref{main2}) is convergent for $\tau > -2$ and $\eta = 1$.
To support such possibility, let us consider an example of the series with an asymptotic of the type (\ref{asymp2}) 
\begin{equation}
  h_k = (-1)^k \frac{\Gamma\Big(\frac{\tau + 4 k + 2}{2}\Big)}{2 \Gamma(k + 1)} u^k\,,~~~u > 0\,.
\label{hk}
\end{equation}
At $\eta = 1$ the substitution of these coefficients into the expansion (\ref{main3}) produces a divergent series.
In opposite, the expansion (\ref{main2}) in this case can be bounded by
\begin{eqnarray}
  \int_{0}^\infty dt\, t^{\tau + 1} e^{-t^2 - \sigma t^4 + |\sigma - u|t^4} < \infty\,.
\label{main4}
\end{eqnarray}
Therefore, the series (\ref{hk}), re-summed in accordance to (\ref{main2}), is convergent.

\subsection{$\gamma$-Regularization}
\label{gammaReg}
The $\eta$-regularization enforce an additional decay of the lattice Boltzmann weight for the large fluctuations
of the fields $\phi_n$. This results in the dumping of the coefficients of the standard perturbation theory
and in the consequent convergence of the series (\ref{main2}) for $|\eta| < \eta_*$.
However, the additional decay is not sufficiently sharp to provide convergence for $\eta_* \leq \eta < 1$,
which is desired, since $\eta = 1$ corresponds to the original theory.
Here we introduce alternative regularization, giving sharper vanishing of the large fluctuations of fields.
We consider the regularized lattice $\phi^4$-model, defined by the action
\begin{equation}
  S_\gamma[\phi] = S[\phi] + \gamma \lVert\phi\rVert^6\,.
\label{vS}
\end{equation}
When $\gamma = 0$ the action $S_\gamma$ coincides with the action of the $\phi^4$-model (\ref{Sphi4}).
For the model defined by (\ref{vS}) one can construct two related expansions.
As always, we demonstrate them on the example of the propagator.
The first one is similar to the standard perturbation theory
and is obtained by expanding the interaction part of the original $\phi^4$-model
into the Taylor series
\begin{equation}
  \langle\phi_i \phi_j\rangle_\gamma = \sum_k \int [d\phi]\, \phi_i\phi_j e^{-\lVert\phi\rVert^2 - \gamma\lVert\phi\rVert^6}
  \frac{\big(-\frac{\lambda}{4!}\sum_n \phi_n^4\big)^k}{k!}\,.
\label{gammaS}
\end{equation}
Rewriting (\ref{gammaS}) in terms of Gaussian integrals analogously to Section \ref{constr} and
introducing the dependence on $\tau$, we get
\begin{eqnarray}
  \langle\phi_i \phi_j\rangle_\gamma
  = \sum_{k = 0}^\infty \frac{2 \int_{0}^\infty dt\, t^{\tau + 4 k + 1} e^{-t^2 - \gamma t^6}}{\Gamma\Big(\frac{\tau + 4 k + 2}{2}\Big)}
  f_k
  \,.
\label{gammaS2}
\end{eqnarray}

The second expansion is the variational series similar to (\ref{main2}).
To derive it, we split the action as
\begin{equation}
  S_\gamma[\phi] = (N[\phi] + \gamma \lVert\phi\rVert^6) + (S[\phi] - N[\phi])\,,
\end{equation}
where $(N[\phi] + \gamma \lVert\phi\rVert^6)$ is treated as the initial approximation.
Then, analogous to previous derivations, we have
\begin{eqnarray}
  \langle\phi_i \phi_j\rangle_\gamma = 
  \sum_{l = 0}^\infty J_\gamma(\tau, l)
  \Bigg[\sum_{k = 0}^{l} C_l^k \sigma^{l - k} \frac{2\, k!\, f_k}{\Gamma\Big(\frac{\tau + 4 k + 2}{2}\Big)}\Bigg]\,,~
\label{vmain}
\end{eqnarray}
where
\begin{eqnarray}
  J_\gamma(\tau, l) = \frac{1}{l!} \int_{0}^\infty dt\, e^{-t^2 - \sigma t^4 - \gamma t^6} t^{\tau + 4l - 1}\,.
\end{eqnarray}

\begin{figure}[H]
  \begin{center}
      \includegraphics[width=0.5\linewidth, angle = 0]{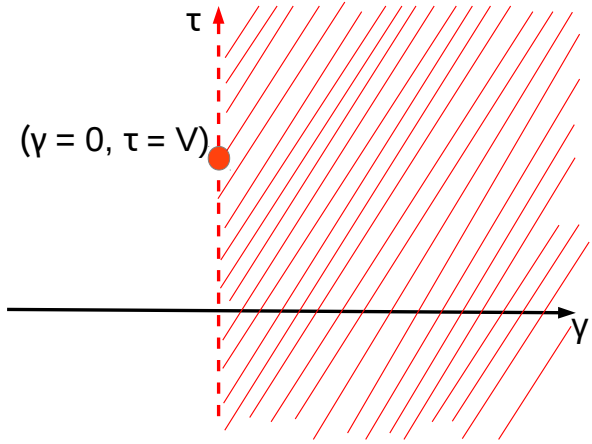}
      \caption{In the marked area the convergence of both series (\ref{gammaS2}) and (\ref{vmain}) is obtained due to the non-zero $\gamma$.
      At $\gamma = 0$ and $\tau = V$, as it follows from the estimate (\ref{est10}), only the series (\ref{vmain}) is convergent.}
      \label{gamalph}
  \end{center}
\end{figure}

The series (\ref{gammaS2}) can be obtained from the series (\ref{vmain})
by changing the order of summations. Consequently, if both of these series
converge absolutely, they converge to the same sum.
The series (\ref{gammaS2}) and (\ref{vmain}) can be bounded by
\begin{eqnarray}
\nonumber
  |\langle\phi_i \phi_j\rangle_\gamma|
  \leq
  \sum_{l = 0}^\infty \big|J_\gamma(\tau, l)\big|
  \Big[\sum_{k = 0}^{l} C_l^k \sigma^{l - k} \frac{2\, k!\, |f_k|}{\Gamma\Big(\frac{\tau + 4 k + 2}{2}\Big)}\Big] \\
  = \sum_{k = 0}^\infty \frac{2 \int_{0}^\infty dt\, t^{\tau + 4 k + 1} e^{-t^2 - \gamma t^6}}{\Gamma\Big(\frac{\tau + 4 k + 2}{2}\Big)}
  |f_k|
  \,.~~~~
\label{vbound}
\end{eqnarray}
At large $k$ the coefficients in front $|f_k|$ in the leading order are determined
by
\begin{equation}
  3 (2\sqrt{3})^{-1-4\beta/3} \gamma^{-2\beta/3} k^{-4\beta/3} (12 e^2 \gamma)^{-2 k/3} k^{-4 k/3}
\label{vasymp}
\end{equation}
with $\beta = (\tau + 2) / 4$.
The bound (\ref{vbound}) converges for any $\tau > -2$, when $\gamma > 0$.
In Fig. \ref{gamalph} we present the area of parameters
$\gamma$ and $\tau$ for which the series (\ref{gammaS2}) and (\ref{vmain}) are convergent.
Similar to the $\eta$-regularization case the convergence of (\ref{vmain})
is better than of the series (\ref{gammaS2}) and bound (\ref{vbound}).

\subsection{Non-perturbative independence on $\tau$ and continuity of the $\gamma$-regularization}
\label{NPInd}
Here we show, that for any $\tau > -2$, there are such $\gamma_* \in \mathbb{R}_+$ and $K_* \in \mathbb{N}$, that
for each $0 < \gamma < \gamma_*$, it is possible to construct convergent series
(\ref{gammaS2}), which approximates the propagator of the lattice $\phi^4$-model
with an arbitrary precision $\delta > 0$:
\begin{equation}
  \Big|
  \sum_{k = 0}^{K_*} \frac{2 \int_{0}^\infty dt\, t^{\tau + 4 k + 1} e^{-t^2 - \gamma t^6}}{\Gamma\Big(\frac{\tau + 4 k + 2}{2}\Big)}
  f_k
  - \langle\phi_i \phi_j\rangle
\Big| < \delta\,.
\label{delta}
\end{equation}
The proof of (\ref{delta}) contains the demonstration of the fact, that in the limit $\gamma \rightarrow 0$ the sum of the
series (\ref{gammaS2}) is independent on $\tau$ non-perturbatively, {\it i.e.}, including all possible non-analytical
contributions. 

Since for $\gamma > 0$ the series (\ref{gammaS2}) is absolutely convergent one can interchange in it the summation
and integration (the SPT-coefficients $f_k$ are integrals), this gives
\begin{equation}
  \langle\phi_i \phi_j\rangle_\gamma =  \int [d\phi]\, \phi_i\phi_j e^{-\lVert\phi\rVert^2} \sum_{k = 0}^\infty h_k(\phi_n, \gamma, \tau)\,,
\label{gammaSD}
\end{equation}
where
\begin{equation}
  h_k(\phi_n, \gamma, \tau) = \frac{\big(-\frac{\lambda}{4!}\sum_n \phi_n^4\big)^k}{k!} \frac{2 \int_{0}^\infty dt\, t^{\tau + 4 k + 1} e^{-t^2 - \gamma t^6}}{\Gamma\Big(\frac{\tau + 4 k + 2}{2}\Big)} \,.
\label{gammaSD2}
\end{equation}
The series $\sum_{k = 0}^\infty h_k(\phi_n, \gamma, \tau)$ converges faster than the series for the exponent and, consequently, is uniformly
convergent for each compact subset of the region of parameters $\bar{\cal A} = \{\phi_n \in \mathbb{R}, \gamma \geq 0, \tau > -2\}$.
In region $\bar{\cal A}$ the functions $h_k(\phi_n, \gamma, \tau)$ are continuous in all parameters, therefore the sum of them is also
continuous. At $\gamma = 0$ each function $h_k(\phi_n, \gamma, \tau)$ is independent on $\tau$ and the propagator (\ref{gammaSD}) coincides
with $\langle\phi_i \phi_j\rangle$, defined by~(\ref{correl}). Hence, (\ref{gammaSD}) is finite at $\gamma = 0$.
The finiteness of (\ref{gammaSD}) for $\gamma > 0$, $\tau > -2$ follows from the convergence of the series
(\ref{gammaS2}). Then, because $\sum_{k = 0}^\infty h_k(\phi_n, \gamma, \tau)$ is continuous in $\bar{\cal A}$
and (\ref{gammaSD}) is finite for $\gamma \geq 0$, $\tau > -2$, it follows, that for any $\widetilde\delta > 0$ 
there is such $\widetilde\gamma$, that
for each $0 < \gamma < \widetilde\gamma$
\begin{equation}
  \Big|
  \langle\phi_i \phi_j\rangle_{\gamma} - \langle\phi_i \phi_j\rangle
\Big| < \widetilde\delta\,.
\label{delta2}
\end{equation}
From the latter inequality and from the convergence of the series (\ref{gammaSD}) we obtain (\ref{delta}).

As it was discussed above the series (\ref{gammaS2}) and (\ref{vmain}) converge to the same quantities, therefore,
all stated here is valid also for (\ref{vmain}).

\subsection{Evaluation of the connected Green's functions}
\label{NPConnect}
In previous parts of the paper we predominantly discussed constructions
of the convergent series for the full Green's functions on the example of the propagator.
The generating functional of the full functions is the partition function $Z[A]$, which is a functional of the external
field $A$. However, the thermodynamical quantities are naturally expressed in terms of the connected
Green's functions, given by derivatives of the $\log(Z[A])$ with respect to the external field $A$.
Within the framework of the standard perturbation theory it is shown, that connected functions are obtained
from the full functions by throwing away all disconnected diagrams of the perturbative expansion, 
see, for instance, \cite{Vasilev}. As it is known from the theory of the combinatorial species
\cite{Leroux} this relation is much more general. If the weights of a generating function 
for a combinatorial weighted species factorize among the connected components of the species, 
then the logarithm of that function is given by the sum over the connected associated species.
In the convergent series (\ref{gammaS2}) as in the standard perturbation theory
the disconnected diagrams are the products of the connected ones, consequently, 
for the computation of the connected functions using the convergent series one has 
to substitute the contributions $f_k$ of all diagrams in each order of SPT by the contributions
only from connected diagrams $\widetilde f_k$. The resulting series is convergent, since the asymptotic of the high orders
of the standard perturbation theory has similar form (\ref{asymp2}) for both full and connected Green's functions.

Let us change $f_k$ by $\widetilde f_k$ in (\ref{vmain}).
Analogously to the relation between (\ref{gammaS2}) and (\ref{vmain}), the interchange of the summations in 
the series (\ref{vmain}) with $\widetilde f_k$ gives the series (\ref{gammaS2}) with $f_k$, changed by $\widetilde f_k$.
Therefore, the series (\ref{vmain}) with $\widetilde f_k$ instead of $f_k$ is also convergent.

\section{Computations and numerical results}
\label{numer}

Here we present the results of the $\langle\phi_n^2\rangle$ operator computations
within the convergent and variational series and compare them with the Borel re-summation and Monte Carlo simulations.
We show the numerical dependences on the variational and regularization parameters $\tau$ and $\gamma$.
The propagator between all lattice cites $\langle\phi_i\phi_j\rangle$ is computed, but we present only the part of the resulting data
representing $\langle\phi_n^2\rangle$, for greater clarity.
All calculations are performed at unit mass $M = 1$ and for the coupling constants $\lambda$ in the region $[0, 10]$.

The computations within the convergent/variational series methods contain two main steps.
The first one is the evaluation of the coefficients of the standard perturbation theory for the propagator $\langle\phi_i\phi_j\rangle$.
For this the connected diagrams of the $\phi^4$-theory are generated using the system 'GRACE' \cite{GRACE}.
The free propagator in diagrams is obtained by the numerical inversion of the $K_{mn}$ matrix (\ref{Kmn}).
The best current results within the $\epsilon$-expansion in the continuum $\phi^4$-theory are obtained in the $6$-loops approximation \cite{Kompaniets}.
We also compute $6$ orders of the lattice SPT, 
since it is reasonable to test the efficiency of the convergent series within the same order of the perturbation theory.
The second step of the calculations using the CS/VS methods is the re-summation of 
the perturbative results in accordance to the formulas (\ref{sum}), (\ref{main}), (\ref{main2}), (\ref{vmain}).

To perform the Borel re-summation procedure, we use the conformal 
mapping for the analytical continuation in the Borel plane.
The conformal mapping can be done if the parameter $a$ from the asymptotic of the high orders of the perturbation
theory (\ref{asymp2}) is known.
We estimate $a$ using the values presented in \cite{Zinn}
for the continuous one and two-dimensional $\phi^4$-model
\begin{eqnarray}
  d = 1 \, , \text{  }~ a = 1 / 8,
\end{eqnarray}
\begin{eqnarray}
  d = 2 \, , \text{  }~ a = 1 / 35.102\dots\,\,.
\label{ad2}
\end{eqnarray}

\begin{figure}[H]
  \begin{center}
  \begin{tabular}{@{}ccc@{}}
    \begin{minipage}[h]{0.5\linewidth}
      \includegraphics[width=\linewidth, angle = 0]{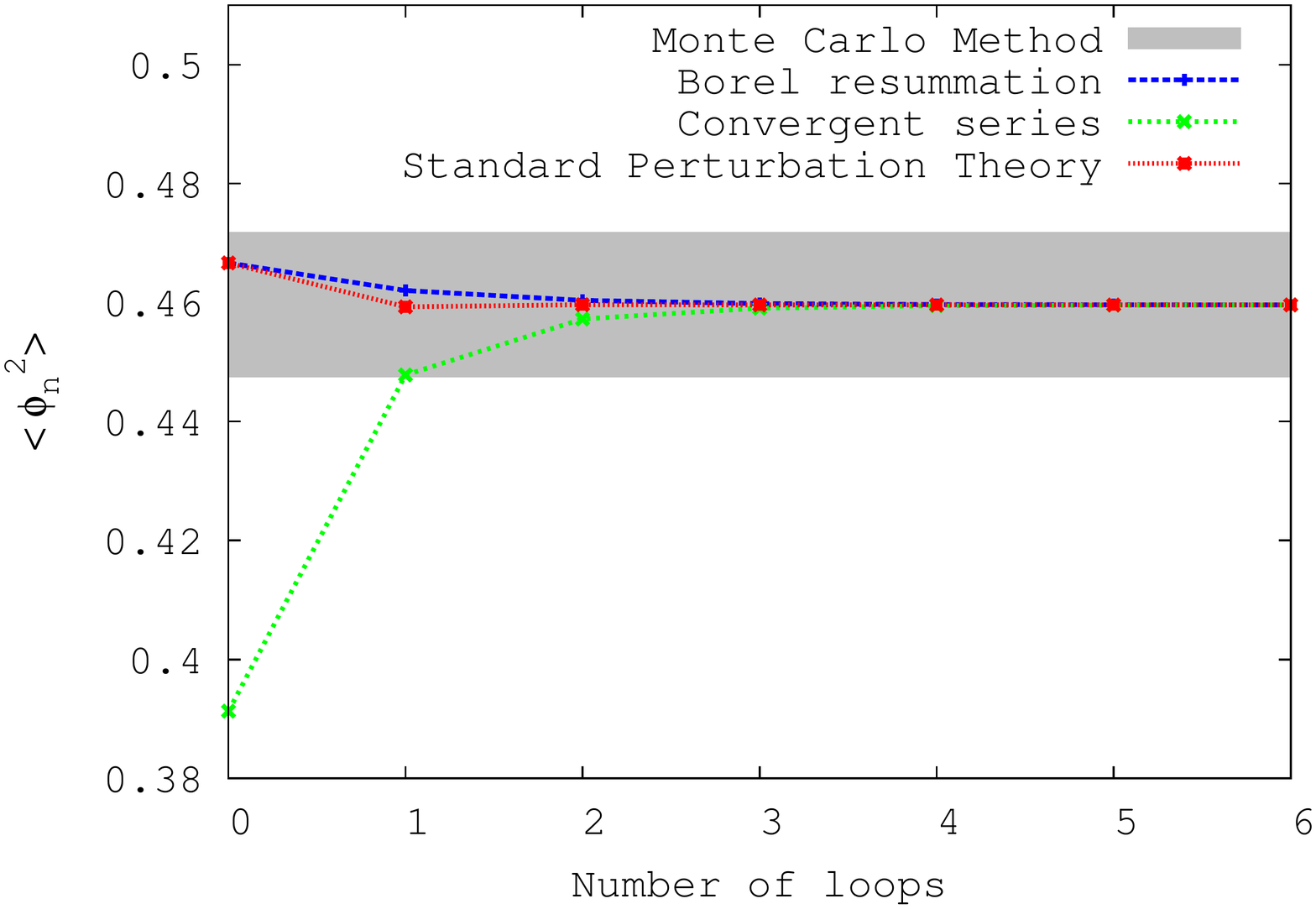}
      \caption{Dependence of $\langle \phi_n^2 \rangle$ on the order of the perturbation theory for the lattice volume $V = 4$ at $\lambda = 0.1$.}
    \label{N4l01}
    \end{minipage}
    ~~~~&
    \begin{minipage}[h]{0.5\linewidth}
      \includegraphics[width=\linewidth, angle = 0]{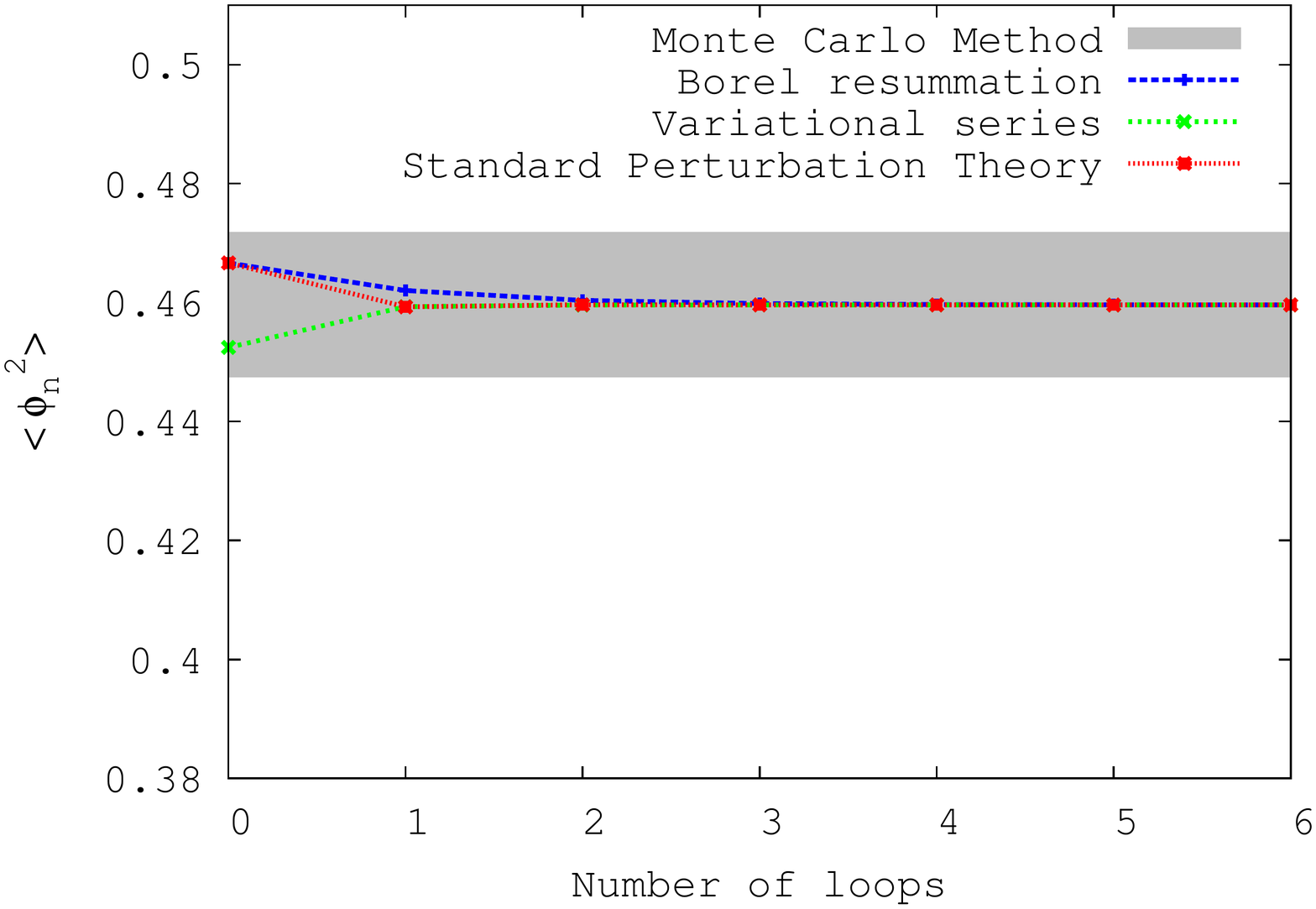}
      \caption{Dependence of $\langle \phi_n^2 \rangle$ on the order of the perturbation theory for the lattice volume $V = 4$, $\tau = 0$ at $\lambda = 0.1$.}
      \label{N4N0l01}
    \end{minipage}
  \end{tabular}
  \end{center}
\end{figure}

\begin{figure}[H]
  \begin{center}
  \begin{tabular}{@{}ccc@{}}
    \begin{minipage}[h]{0.5\linewidth}
      \includegraphics[width=\linewidth, angle = 0]{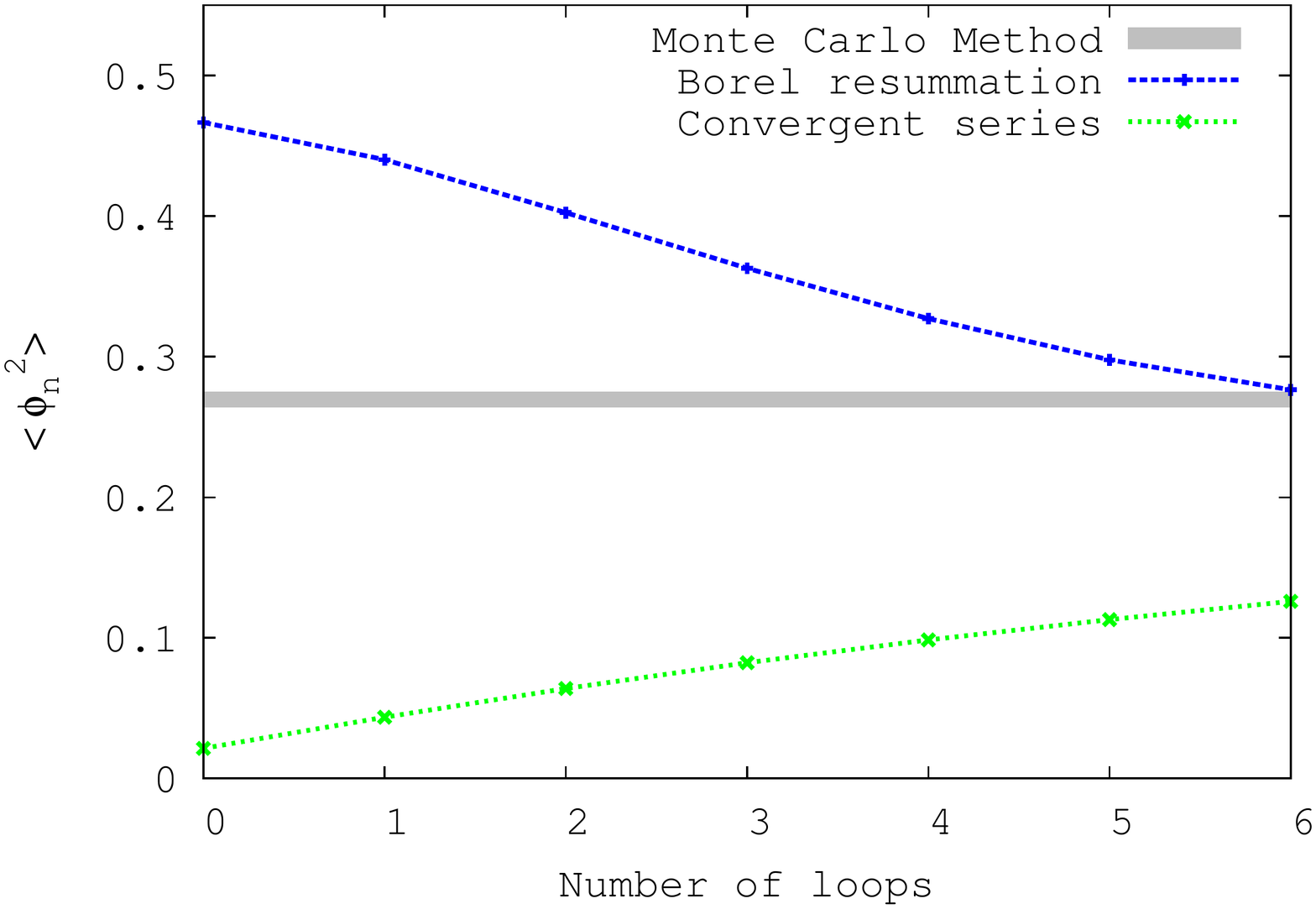}
      \caption{Dependence of $\langle \phi_n^2 \rangle$ on the order of the perturbation theory for the lattice volume $V = 4$ at $\lambda = 10$.}
      \label{N4l10}
    \end{minipage}
    ~~~~&
    \begin{minipage}[h]{0.5\linewidth}
      \includegraphics[width=\linewidth, angle = 0]{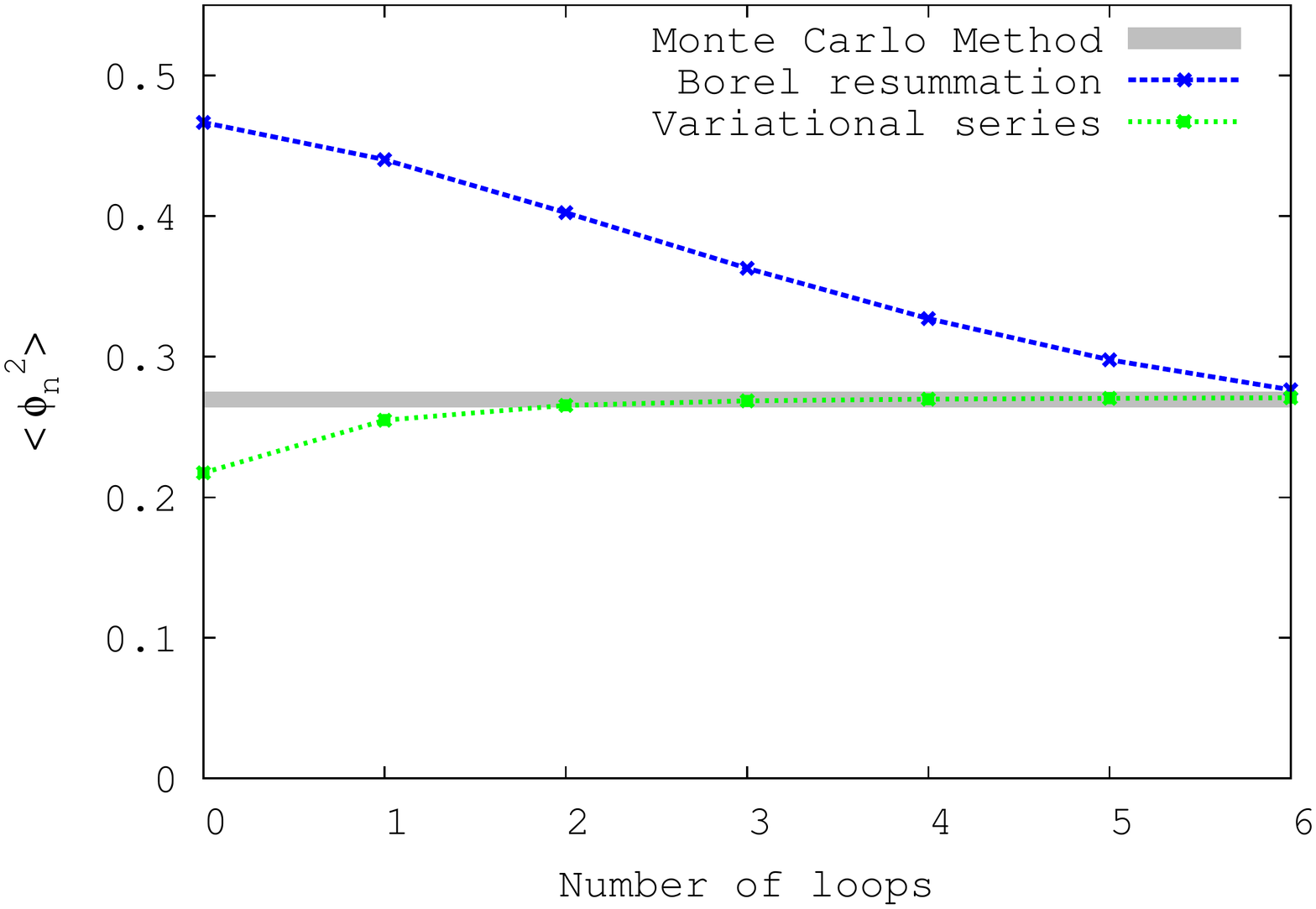}
      \caption{Dependence of $\langle \phi_n^2 \rangle$ on the order of the perturbation theory for the lattice volume $V = 4$, $\tau = 0$ at $\lambda = 10$.}
      \label{N4N0l10}
    \end{minipage}
  \end{tabular}
  \end{center}
\end{figure}

The results obtained on the lattice with $V = 4$ sites are the following.
In Figs. $\ref{N4l01}$ and $\ref{N4l10}$ we present the operator 
$\langle \phi_n^2 \rangle$ computed with CS and other methods depending on the number of known orders of the
perturbation theory (from $0$ to $6$ loops) at coupling constants $\lambda = 0.1$ and $\lambda = 10$ respectively.
At small value of the coupling constant $\lambda = 0.1$ one observes an agreement between all methods including the standard perturbation theory.
When $\lambda = 10$, the result is qualitatively different. The convergent series is out of the Monte Carlo error bars, but
the Borel re-summation method is in the agreement with the Monte Carlo data. 
The values of the standard perturbation theory series are omitted because of the strong divergence.

In Figs. $\ref{N4N0l01}$ and $\ref{N4N0l10}$ we show the results obtained with the variational
series and other methods at coupling constants $\lambda = 0.1$ and $\lambda = 10$ respectively.
The variational series exhibits remarkable agreement with the Monte Carlo data and converges
even faster than the Borel re-summation.

\begin{figure}[H]
  \begin{center}
  \begin{tabular}{@{}ccc@{}}
    \begin{minipage}[h]{0.5\linewidth}
      \includegraphics[width=\linewidth, angle = 0]{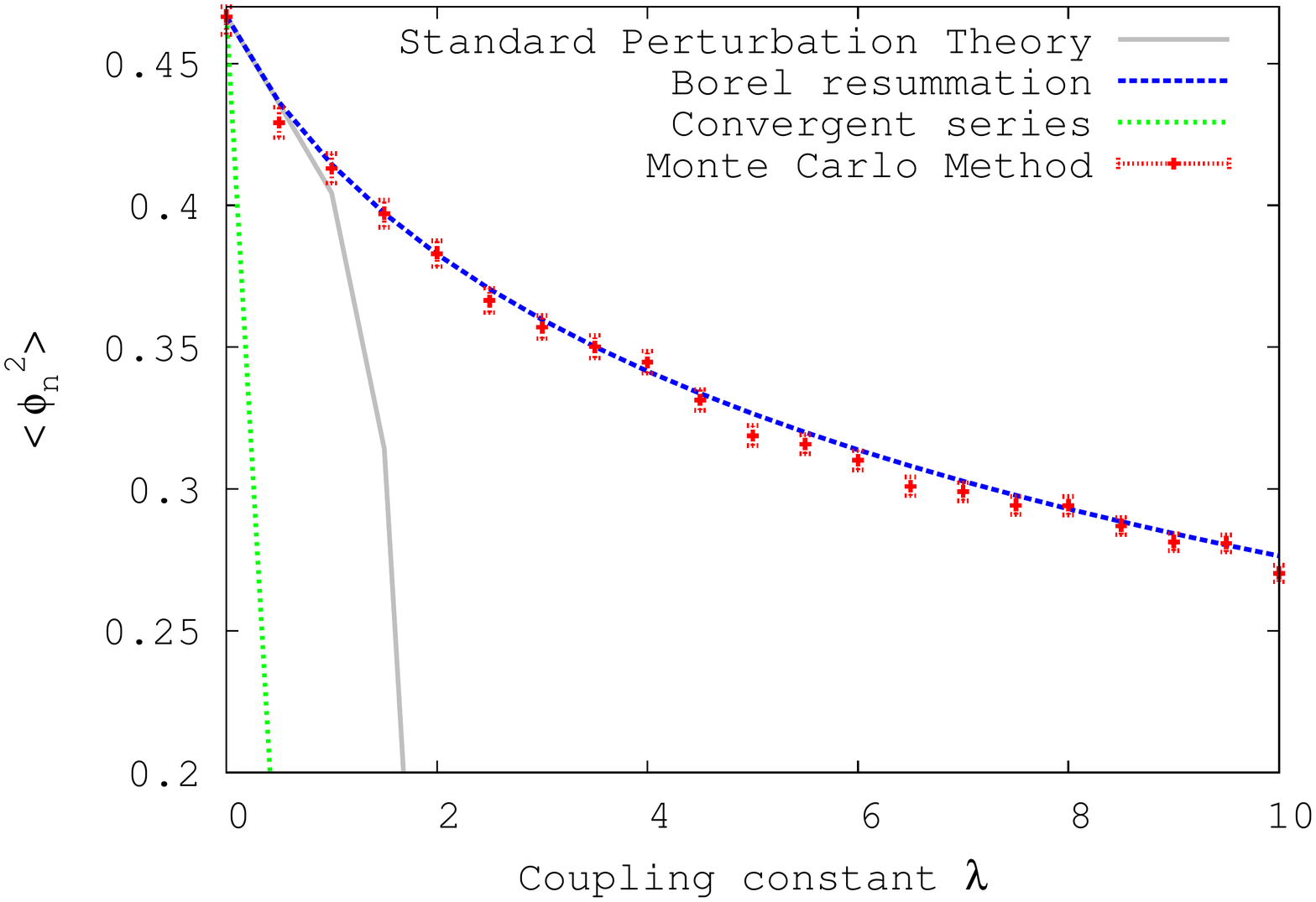}
      \caption{Dependence of the operator $\langle \phi_n^2 \rangle$ on coupling constant for the lattice volume $V = 4$.}
      \label{N4l0-10}
    \end{minipage}
    ~~~~&
    \begin{minipage}[h]{0.5\linewidth}
      \includegraphics[width=\linewidth, angle = 0]{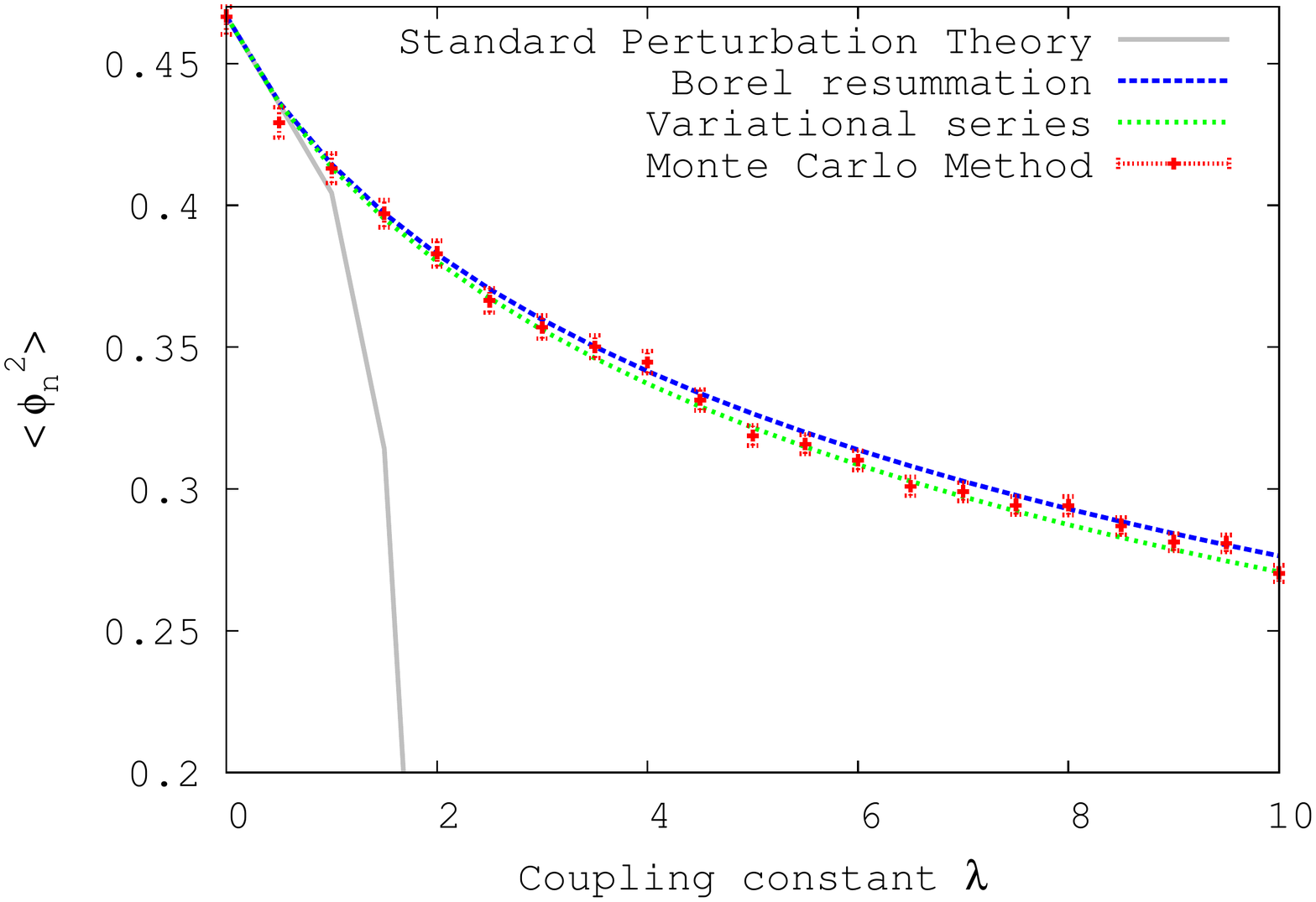}
      \caption{Dependence of the operator $\langle \phi_n^2 \rangle$ on coupling constant for the lattice volume $V = 4$, $\tau = 0$.}
      \label{N4N0l0-10}
    \end{minipage}
  \end{tabular}
  \end{center}
\end{figure}

In Figs. $\ref{N4l0-10}$ and $\ref{N4N0l0-10}$ we demonstrate the dependence of $\langle \phi_n^2 \rangle$ 
on the coupling constant $\lambda$ for different methods with $V = 4$. The computations carried according to
 VS agree with the Monte Carlo results with the precision of the one standard deviation.

\begin{figure}[H]
  \begin{center}
  \begin{tabular}{@{}ccc@{}}
    \begin{minipage}[h]{0.5\linewidth}
      \includegraphics[width=\linewidth, angle = 0]{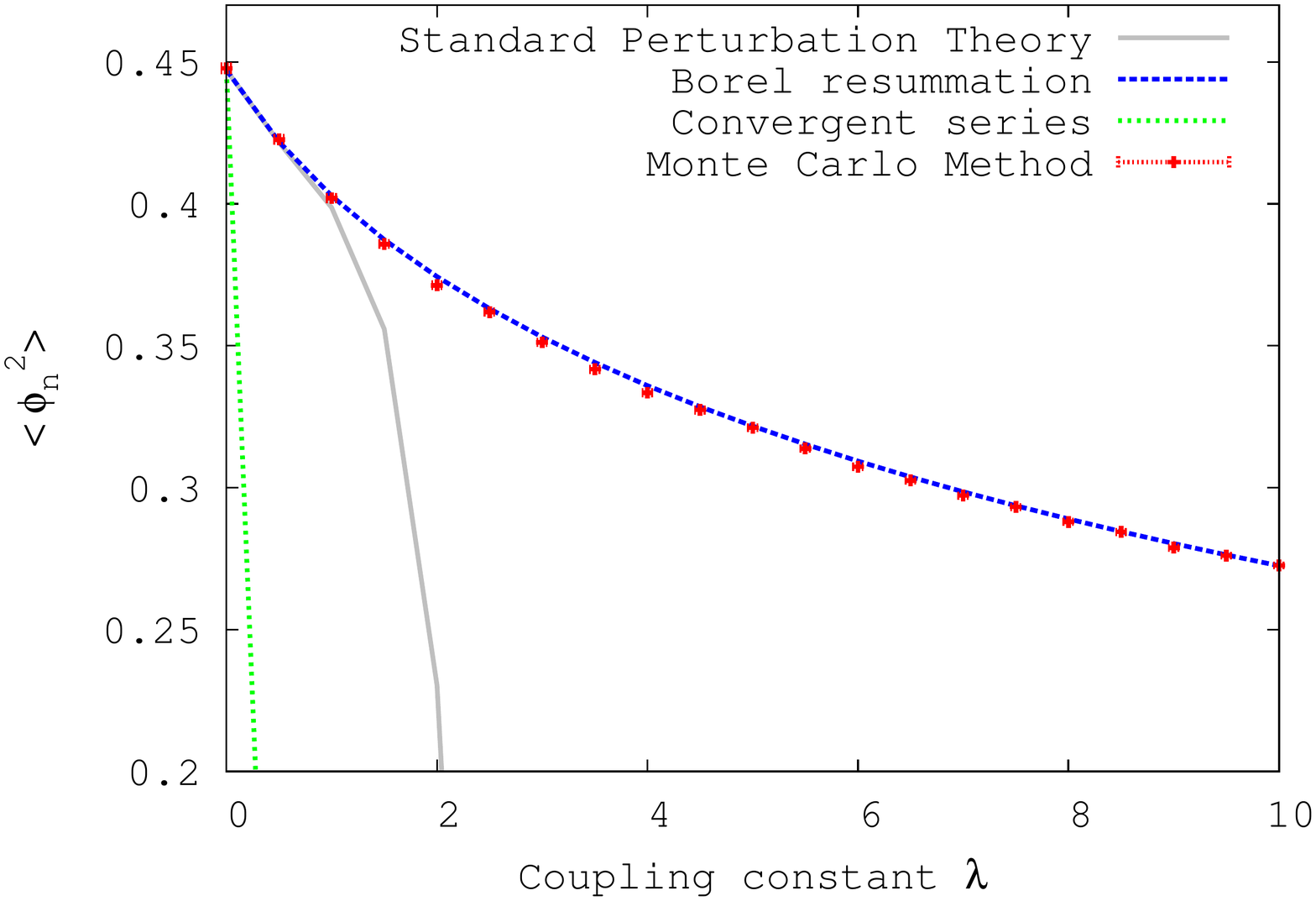}
      \caption{Dependence of the operator $\langle \phi_n^2 \rangle$ on coupling constant for the lattice volume $V = 64$.}
      \label{N64l0-10}
    \end{minipage}
    ~~~~&
    \begin{minipage}[h]{0.5\linewidth}
      \includegraphics[width=\linewidth, angle = 0]{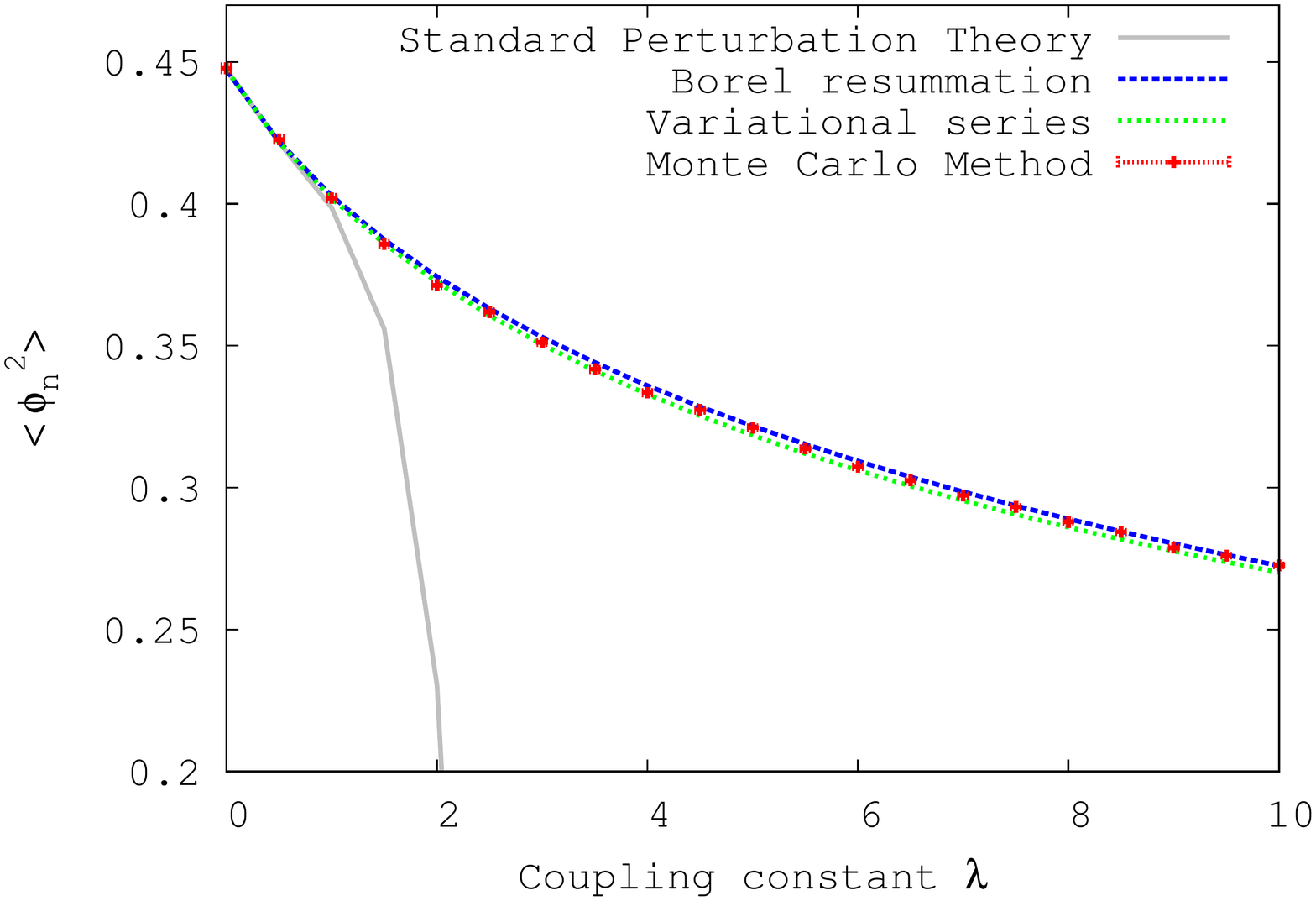}
      \caption{Dependence of the operator $\langle \phi_n^2 \rangle$ on coupling constant for the lattice volume $V = 64$, $\tau = 0$.}
      \label{N64N0l0-10}
    \end{minipage}
  \end{tabular}
  \end{center}
\end{figure}

\begin{figure}[H]
  \noindent\centering{\includegraphics[width=0.55\linewidth, angle = 0]{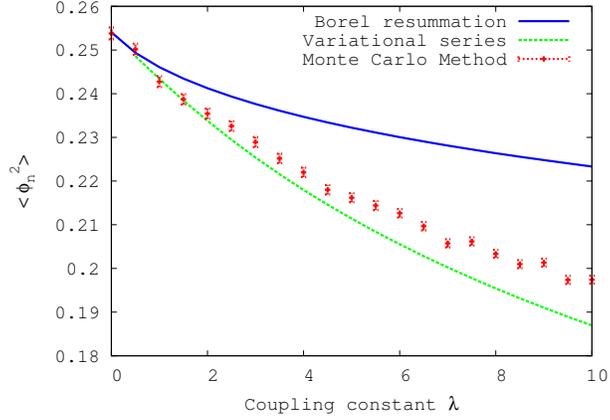}}
  \caption{Dependence of the operator $\langle \phi_n^2 \rangle$ on coupling constant in the two-dimensional case $V = 8\times 8$, $\tau = 0$.}
\label{N8x8N0l0-10}
\end{figure}

\begin{figure}[H]
  \noindent\centering{\includegraphics[width=0.55\linewidth, angle = 0]{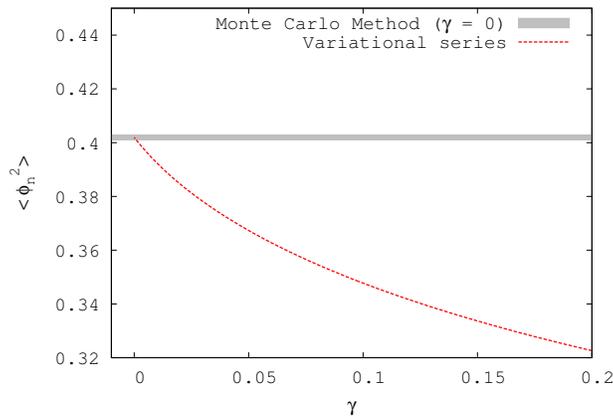}}
  \caption{Dependence of the operator $\langle \phi_n^2 \rangle$, computed by the variational series, 
  on the regularization parameter $\gamma$, $\lambda = 1$, $V = 64$, $\tau = 0$.}
\label{N0l1gamma}
\end{figure}

The results obtained on the one-dimensional lattice with $V = 64$ are qualitatively similar. 
In Fig. $\ref{N64l0-10}$ 
we show the computations performed with the convergent series method in comparison with other methods.
Even at small coupling constants the convergent series does not agree with the Monte Carlo. 
The Borel re-summation is in agreement with the Monte Carlo data in the whole region $\lambda \in [0, 10]$. 
However, the results obtained within VS are significantly different from the
results of the simple convergent series. Corresponding computations are presented in Fig. 
$\ref{N64N0l0-10}$. The variational series matches the Monte Carlo data with the precision of 
the one standard deviation again!
In Fig. $\ref{N8x8N0l0-10}$ we present the dependence of $\langle \phi_n^2 \rangle$ on $\lambda$ 
for the two-dimensional lattice with $V = 8 \times 8$, calculated with the VS method, $\tau = 0$. 
The VS method gives the results which are close to the Monte Carlo simulations. The strong deviation
of the Borel re-summation can be caused by the not precise estimate of the parameter $a$ in the asymptotic 
of the high orders of the perturbation theory (\ref{ad2}).
In Section \ref{NPInd} we have proved the continuity of the series (\ref{vmain}) with respect
to the parameter $\gamma \geq 0$, in Fig. \ref{N0l1gamma} we demonstrate it for $\lambda = 1$.

\begin{figure}[H]
  \begin{center}
  \begin{tabular}{@{}ccc@{}}
    \begin{minipage}[h]{0.5\linewidth}
      \includegraphics[width=\linewidth, angle = 0]{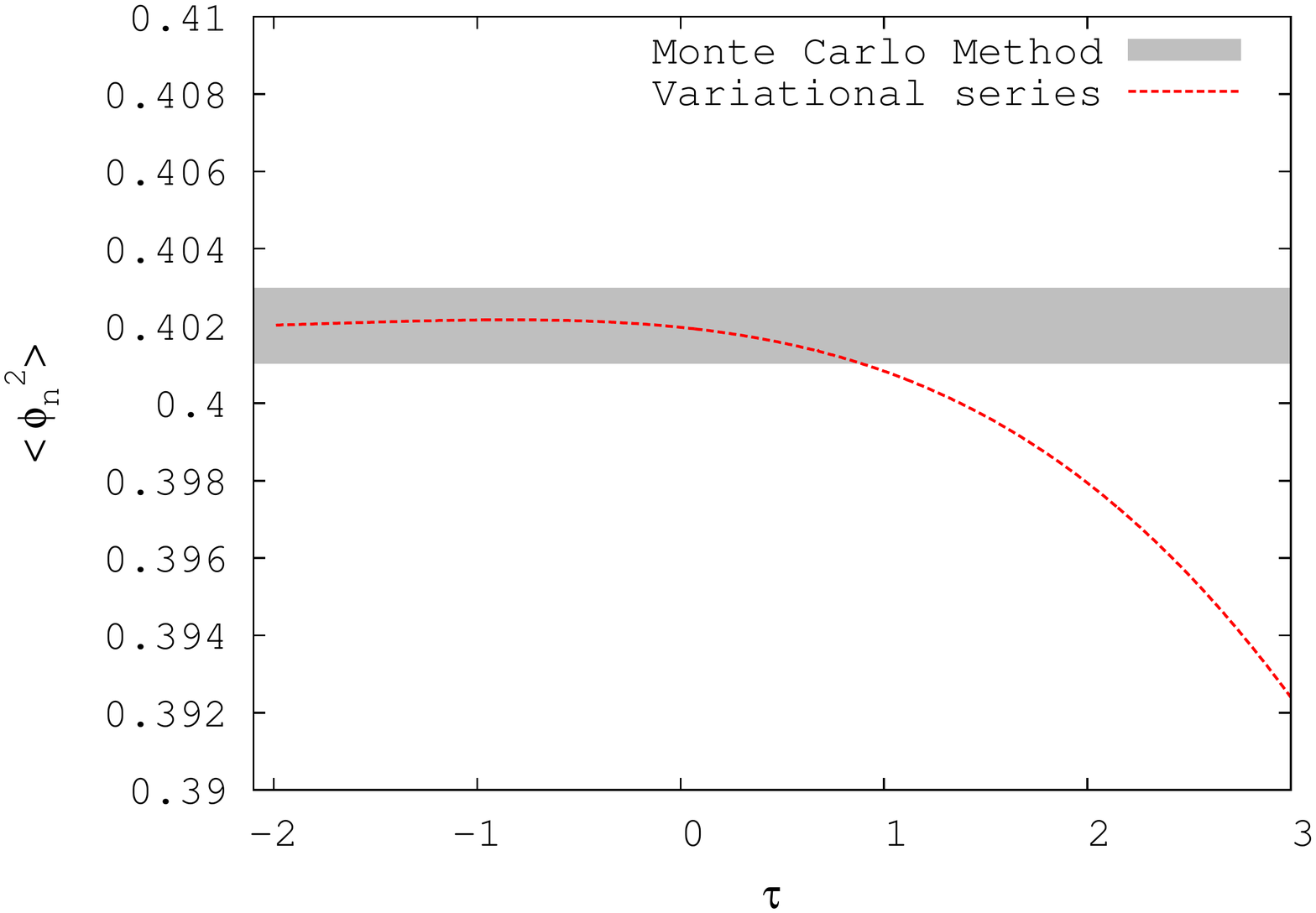}
      \caption{Dependence of the operator $\langle \phi_n^2 \rangle$ on the variational parameter $\tau$, $\lambda = 1$ for the lattice volume $V = 64$.}
      \label{NXl1}
    \end{minipage}
    ~~~~&
    \begin{minipage}[h]{0.5\linewidth}
      \includegraphics[width=\linewidth, angle = 0]{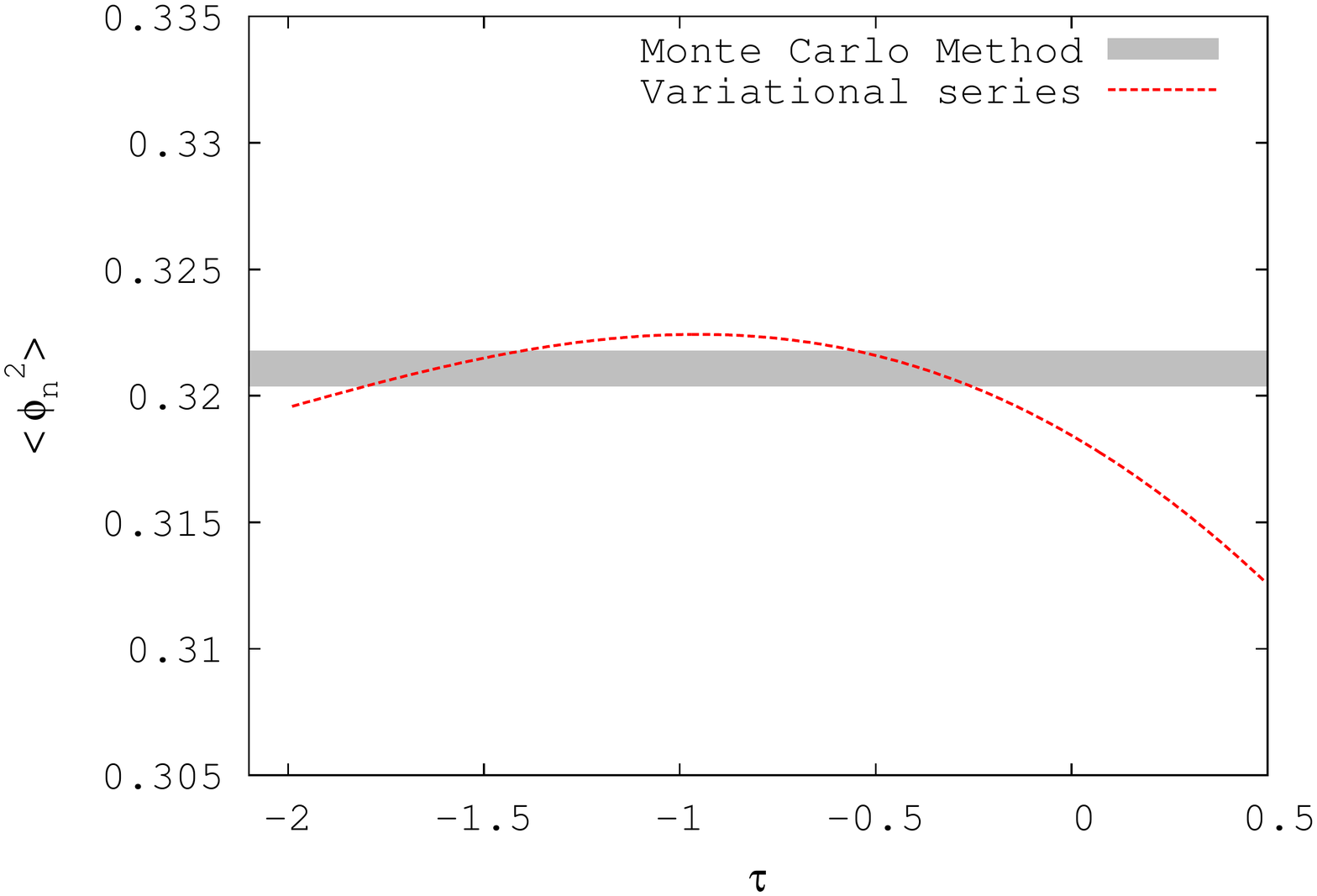}
      \caption{Dependence of the operator $\langle \phi_n^2 \rangle$ on the variational parameter $\tau$, $\lambda = 5$ for the lattice volume $V = 64$.}
      \label{NXl5}
    \end{minipage}
  \end{tabular}
  \end{center}
\end{figure}

In the limit $\gamma \rightarrow 0$ the dependence on $\tau$ of the sum (\ref{vmain})
has to disappear. In the real calculations one always has only a finite amount
of terms of (\ref{vmain}). In Figs. \ref{NXl1} and \ref{NXl5} we present the dependence
on $\tau$ of the $6$-th order approximation of the variational series (\ref{main2}) for the 
coupling constants $\lambda = 1$ and $\lambda = 5$ respectively. The optimal values of the parameter $\tau$,
giving the matching with the Monte Carlo mean value,
are $\tau_{\lambda = 1} \simeq -0.1$,
$\tau_{\lambda = 5} \simeq \{-1.62; -0.38\}$. Without comparison with the Monte Carlo simulations one can apply
the principle of the smallest contribution of the last term of the series, it
gives uniform optimal $\tau_\lambda \simeq -0.086$. In the main part of our computations we use $\tau = 0$,
what corresponds in the continuum limit to the utilization of the dimensional regularization \cite{Leibbrandt}.

\section{Conclusions}
\label{concl}
We have presented the construction of the convergent series for the lattice models with the even degree
polynomial interaction and investigated it in details on the example of the lattice $\phi^4$-model.
We have proved, that CS is a re-summation method. The latter fact supports the resurgence idea, that the
non-perturbative physics can be expressed via the coefficients of the standard perturbation theory.
The initial approximation
of the convergent series is a non-Gaussian interacting non-local model, hence, CS automatically
takes into account such non-analytical contributions as $e^{-\frac{1}{\lambda}}$.

We have observed an internal symmetry of the CS method and, using it, developed the variational
series.
The numerical values of the operator $\langle \phi_n^2 \rangle$ calculated using the CS and VS methods
with $6$ orders of the standard perturbation theory
were compared with the Borel re-summation and Monte Carlo method. 
The comparison revealed, that for the small lattices and for the small coupling constants
the convergent series exhibits the agreement with the Monte Carlo and Borel re-summation, but deviates
from them for larger lattice volumes and/or for larger coupling constants. 
The variational series method, applied to the one- and two-dimensional cases, agrees
with other methods for the wide range of the lattice
volumes and coupling constants and converges even faster to the Monte Carlo results than the Borel re-summation.

To study the convergence and correctness of the variational series from the analytical point of view,
we have considered two regularizations of the lattice $\phi^4$-model. In both cases we have constructed for the variational series
the upper bounds with finite radii of convergence in terms of the regularization parameters.
For the $\gamma$-regularization the border of the convergence region approaches $\gamma = 0$, which corresponds to the
non-regularized model. 
We also have shown, that original model can be approximated by the
the $\gamma$-regularization with any arbitrary precision and 
it is always possible to construct two related convergent expansions for this regularization.
One of them is precisely the variational series, derived for the regularized model.
The convergence properties of VS, depending on the regularization parameters are summarized in Figs. \ref{epsalph}
and \ref{gamalph}. Using this information and the fact, that the convergence of the series (\ref{vmain}) and (\ref{main2}) should be better,
than the convergence of their bounds, we conjecture, that the series (\ref{main2}) is convergent
for any $\tau > -2$.

The convergence of (\ref{main2}) independently on the values of $\tau$ and the independence on $\tau$ of its sum at $\gamma \rightarrow 0$ allows
to consider any finite $\tau > -2$ even for very large (infinite) volumes $V$. The diagrams of the standard perturbation theory
can be easily computed at infinite volume. This provides a way for taking the infinite volume limit within the CS/VS methods.
The VS computations at $\tau = 0$ work equally well for the lattice volumes more than $10$ times different, see Fig. \ref{N4N0l0-10}
and \ref{N64N0l0-10}, it gives a numerical evidence of the possibility to perform the infinite volume limit.

It is important to note, that the applicability of the convergent/variational series is not based on any
criteria of the kind of the Borel summability of the original perturbation theory.
The presented construction can be generalized to the fermionic lattice models by employing bosonization.
For instance, an application of the method \cite{Belokurov1, Belokurov2, Belokurov3, BelSolSha97, BelSolSha99} 
to the bosonized fermionic model, a model of lattice QED, was proposed in \cite{Sazonov2014}.
The problem of the bosonizations of the complex actions has been recently solved in \cite{nonHBos}.
Therefore, the convergent series, which is based only on the perturbative computations, can provide a way for the
bypassing the sign problem in bosonic and fermionic models.

{\bf Acknowledgements}
We acknowledge Vladimir V. Belokurov, Eugeny T. Shavgulidze, Vincent Rivasseau,
Vladimir Yu. Lotoreychik for the discussions and Nikolay M. Gulitskiy for reading
the manuscript.
The work of A.S. Ivanov was supported by the Russian Science
Foundation grant 14-22-00161. V.K. Sazonov acknowledges the  Austrian  Science
Fund FWF, Grant. Nr. I 1452-N27. 

\label{Bibliography} 
\bibliographystyle{unsrt}
\bibliography{bibliography}

\end{document}